\documentclass[hyper]{JHEP} 
\title{Remark about string field for general configuration
of $N$ D-instantons}
 \author{by J. Kluso\v{n}\\ 	 Department of Theoretical Physics and Astrophysics\\                    Faculty of Science, Masaryk University\\ Kotl\'{a}\v{r}sk\'{a} 2, 611 37, Brno\\ Czech Republic\\ 	E-mail: \email{klu@physics.muni.cz}}   \preprint{\hepth{0109049}}  			  	
 \abstract{In this paper we would like to
suggest matrix form of  the string
field  for any configuration of N D-instantons
in bosonic string field theory.}
    \def\tr{\mathrm{Tr}}

\def\ket #1{\left|#1\right>}

\begin{document}
\section{Introduction}\label{first}

Renewed attention has been paid to
Witten's cubic bosonic open string field
theory \cite{WittenSFT}, following Sen's
conjectures that this formalism can be
used to give analytic  description of
D25-brane decay in bosonic string theory
\cite{SenT} (For  review and extensive
list of references, see \cite{Ohmori}.)
It seems to be possible
 that string field theory could
give very interesting information about
nonperturbative nature of string theory and
consequently about M theory.
For that  reason it seems to be interesting
to study the relation between string field theory
and matrix theory \cite{Banks}, which is the most successful
nonperturbative definition of M theory.
For example,  recent progress
in the vacuum string field theory
\cite{SenV1,Hata,Taylor,Feng,Feng1,
SenV1,SenV3,Gross1,Kawano,SenV4,Muki,
David,SenV5,Gross2,Matsuo,Okuyama} suggests that
the string field theory 
could be useful for better understanding of the basic
fabric of the  string theory.

In order to find  relation between matrix theory
and string field theory, it
would be perhaps useful to study the nonabelian
extension of 
 string field theory as well. As was stressed
in the original paper \cite{WittenSFT}, nonabelian
extension of string field theory can be very easily
implemented into its formalism
 by introducing Chan-Paton factors
for various fields in the string field theory action
and including the trace over these indices.
 In modern language this
configuration corresponds to $N$ coincident 
D25-branes.  

In the previous paper \cite{KlusonMSFT}
we have proposed a generalised form
of the string field theory action that was
suitable for the
description of general configuration
of D-instantons. We have formulated this
theory in a pure abstract form following the
seminal paper \cite{WittenSFT}.
In order to support further our proposal, we
think that it would be desirable to have an
alternative formulation of the generalised matrix string field
theory action which would allow us to perform
 more detailed
calculation. In particular, it would be
nice to have a matrix generalisation
of the action written in the conformal field
theory (CFT) language
\cite{Leclair1,Leclair2}. In this
paper  we
suggest a possible
form of the matrix  valued string fields
that will be building blocks for the matrix
CFT formulation of the string field action.
We present a compact form of 
this matrix valued string field. We
will study its operator product expansion
(OPE) with the open string stress energy tensor and
we will show that in order that any general component
of the string field to have a  
well defined conformal dimension
the
background configuration of $N$ D-instantons
(In this paper we will discuss D-instantons only,
the extension to  Dp-branes of any dimension is
trivial.) must obey one particular condition
that can be interpreted as a requirement that the
background configuration of D-instantons
is a solution of the equation of motion  arising
from the low energy action for the
D-instanton matrix model. In our opinion this situation
is similar with the fact
that consistent string field theory should be
formulated  using conformal field theory
that forces the background field to obey the
equation of motion. 

Then we extend our analysis to the case
of infinitely many D-instantons and we show
that well known nonabelian configuration 
can be  very easily included in our formalism.
In particular, we find such a matrix form of
the string field and hence vertex operators
that  precisely corresponds to the string field
theory formulated around a noncommutative 
D-brane background \cite{WittenNG,WittenNG1}.

In conclusion we outline our  results and suggest
extension of this  work. In particular, it
will be clear from this paper that the extension
of our approach to the supersymmetric
case can be very easily
performed.

\section{String field theory in the CFT formalism}
In this section we review basic facts about
bosonic string field theory, following
mainly \cite{Ohmori,SenT}. Gauge invariant
string field theory is described with the
full Hilbert space of the first quantized open
string including $b,c $ ghost fields subject
to the condition that the states must
carry ghost number one, where $b$ has ghost
number $-1$, $c$ has ghost number $1$ and
$SL(2,C)$ invariant vacuum $\ket{0}$
carries ghost number $0$. We denote
$\mathcal{H}$ the subspace of the full Hilbert
space carrying ghost number $1$.  Any state
in $\mathcal{H}$ will be denoted as $\ket{\Phi}$
and corresponding vertex operator $\Phi(x)$
is the  vertex operator that  creates state $\ket{\Phi}$
out of the vacuum state $\ket{0}$
\begin{equation}
\ket{\Phi}=\Phi (x)\ket{0} \ .
\end{equation}
Since we are dealing with open string theory,
the vertex operators should be put on the boundary
of the world-sheet. 

The open string field theory action has a form
\begin{equation}
S=\frac{1}{g_0^2}\left(
\frac{1}{2\alpha'}\left<I\circ \Phi (0) Q_B \Phi (0)\right>+
\frac{1}{3}\left<f_1\circ \Phi(0) f_2\circ \Phi(0)
f_3\circ \Phi(0)\right>\right) \ ,
\end{equation}
where $g_0$ is open string coupling constant, $Q_B$
is BRST operator and $<>$ denotes correlation function
in the combined matter ghost conformal field theory.
$I,f_1, f_2, f_3$ are conformal mapping  exact
form of which is reviewed in \cite{Ohmori} and $f_i\circ\Phi(0)$ denotes
the conformal transformation of $\Phi(0)$ by $f_i$. For
example, for $\Phi$ a primary field of dimension $h$,
then $f_i\circ \Phi(0)=(f'_i(0))^h\Phi(f_i(0))$. 

We can expand any state $\ket{\Phi}\in \mathcal{H}$ as
\begin{equation}\label{fieldinfo}
\ket{\Phi}=\Phi(0)\ket{0}=(\phi(y)+A_{\mu}(y)\alpha^{\mu}_{-1}+
B_{\mu\nu}(y)\alpha^{\mu}_{-1}\alpha^{\nu}_{-1}+\dots)c_1
\ket{0}=\sum_{\alpha} \phi^{\alpha}(y)\Phi_{\alpha}(0)\ket{0} \ ,
\end{equation}
where coefficients $\phi^{\alpha}(y)$ in front of basis states 
$\ket{\Phi_{\alpha}}$ of $\mathcal{H}$
depend on the centre-of-mass state 
coordinate $y$ and where index $\alpha$
labels all possible vertex operators of
ghost number one. As we think of the coefficient functions
$\phi_{\alpha}(y)$ as space-time particle fields, we
call $\ket{\Phi}$  as a string field \cite{Ohmori}. The vertex
operator $\Phi(z)$ defined above is also called as a string
field. 

The previous action describes string field theory living
on one single D25-brane. In order
to describe a configuration of $N$ coincident D25-branes we
equip the open string with Chan-Paton 
degrees of freedom so that coefficient functions become
matrix valued and so $\ket{\Phi}$. In the following
we restrict to the case of $N$ D-instantons where
strings obey Dirichlet boundary conditions in all dimensions
and where coefficient functions are  $N\times
N$ matrices without any dependence on $y$.
 We will write such
a string field as $\ket{\hat{\Phi}}$
and call it mostly in the text as a \emph{"Matrix valued
string field"} keeping in mind that this is $N\times N$ matrix
where each particular component $\ket{\hat{\Phi}}_{ij}$ corresponds
to the string field that describes the state of the
 string connecting the i-th D-instanton with the j-th D-instanton.

As we claimed in the introduction, it would be interesting to
have a formulation of the string field action for any configuration
of D-instantons. While some progress in this  direction has
been made in \cite{KlusonMSFT}, we would like to find
such a formulation of the action based on the CFT description. As the
first step in searching such a string field theory action we propose generalised matrix
valued vertex operators carrying CP factors that describe any
configuration of  D-instantons. We will discuss this approach in
the next section.

\section{String fields for $N$ D-instantons}

We propose the form of the matrix valued string field which
in our opinion provides description of the 
general configuration of $N$ D-instantons  in the
bosonic  string field theory.
 D-instantons are characterised by the
strings having Dirichlet boundary conditions in all 
dimensions $y^I,
I=1,\dots, 26$. Let us consider the situation with
$N$ D-instantons placed in  general positions. 
This configuration is described 
 with the matrices
\begin{equation}\label{clas}
Y^I=\left(\begin{array}{cccc}
y^I_1 & 0 & \dots & 0 \\
0 & y^I_2 & \dots & 0 \\
\dots & \dots & \dots & \dots \\
0 & \dots & 0 & y^I_N \\ \end{array}\right) , \ I=1,\dots, 26 \ ,
\end{equation}
where $y_i^I$ labels coordinate of i-th D-instanton. 
Moreover, the  configuration (\ref{clas}) corresponds
to the solution of the  equation of motion of the low
energy matrix model effective action and as we will
see, some consistency requirements that will be
posed on the matrix valued string fields also
imply that the background configuration of $N$ D-instantons
(\ref{clas}) should have this form. 

It is well known that the string stretching
from the i-th D-instanton to the j-th D-instanton has an energy
proportional to the distance between these two branes.
More precisely,   vertex operator describing
the ground state 
 of the string going from the i-th D-instanton
to the j-th D-instanton is given by
\footnote{ Because we implicitly
presume that $U$ is normal ordered we will not write  symbol of
the normal ordering $:$. For simplicity, we will
also consider the dependence of the world-sheet
fields $X^I(z)$ on the holomorphic coordinate $z$ only. }
\begin{eqnarray}
\ket{{ij}}\equiv u_{ij}(z=0)c(0)\ket{0} 
\equiv c(0)
\exp\left(\frac{i(y^I_i-y^I_j)}{2\pi\alpha'}
g_{IJ}X^J(0)\right)\ket{0} \ , i, j =1,\dots, N \ ,
 \nonumber \\
\end{eqnarray}
with $\ket{0}$ being the $SL(2,C)$ invariant vacuum state.
In the previous expression $g_{IJ}$ is a flat closed string metric 
$g_{IJ}=\delta_{IJ}$ with signature $(+,\dots,+)$.  
Let us consider some state of ghost number one from the
 first quantized Hilbert space of the open string that
does not depend on the zero mode
of $X^I(z)$ which means that  $\Phi_{\alpha}=\Phi_{\alpha}
(\partial X, c,b)$
 commutes with $u_{ij}$ given above. The index
$\alpha$ labels all possible vertex operators of ghost number
one.  Then any string
field 
corresponding to the string going from the i-th D-instanton
to the j-th D-instanton can be written in the similar form
as in (\ref{fieldinfo})
\begin{equation}
\ket{\hat{\Phi}}_{ij}=\sum_{\alpha} 
 A_{ij}^{\alpha}\Phi(0)_{\alpha} u_{ij}(0)\ket{0} \ , 
\end{equation}
where $(A)_{ij}^{\alpha}$ is analogue of $\phi^{\alpha}(y)$ in
(\ref{fieldinfo}).  Roughly
speaking, matrix $A^{\alpha} \in U(N)$ contains information 
which string  from the collection of all possible $N^2$ strings
of the system of $N$ D-instantons 
(or more precisely,
with which amplitude of probability)
 is excited in given state 
characterised by the world-sheet operator $\Phi_{\alpha}(z)$. In the
 following we restrict
ourselves to  one particular CFT operator $\Phi_{\alpha}(z)$
 and its corresponding $A^{\alpha}$. For that reason we omit the
index $\alpha$  in our formulas. In spite of this fact we will
still call $\hat{\Phi}$   a string field since it describes the
whole system.

{}From the previous analysis it is clear that any string field is $N\times N$ matrix 
that in   more detailed description has a  form
\begin{equation}\label{fi}
\hat{\Phi}(0)=\left(\begin{array}{cccc}
A_{11} u_{11}(0) &   A_{12}u_{12}(0) & \dots &
A_{1N}u_{1N}(0) \\
A_{21} u_{21}(0) & A_{22}u_{22}(0)
 & \dots & A_{2N}u_{2N}(0) \\
\dots & \dots & \dots & \dots \\
A_{N1}u_{N1}(0) & \dots & A_{N,N-1}u_{N,N-1}(0) &
A_{NN}u_{NN}(0) \\ \end{array}\right) 
\times \Phi(0) \ . 
\end{equation}
We would like to argue that this expression can be written in 
more symmetric form.  Let us define  $N\times N$ matrix-
 operator  that is a function of the matrices $Y^I$ and
the world-sheet fields $X^I(z)$ 
\begin{equation}\label{U}
U(z)(\cdot \ )=\exp\left(\frac{i}{2\pi
\alpha'}[Y^I, \cdot \  ] g_{IJ}X^J(z)\right)  \ ,
\end{equation}
where  definition of its action on any $N\times N$ matrix
 will
be  given below.
Our proposal is that the generalised
matrix valued string field  can be written as
\begin{equation}\label{nphi}
\hat{\Phi}(z)= U(z)(A)\Phi(z) \ .
\end{equation}
We will show that for the background given
(\ref{clas}) the operator (\ref{nphi}) reduces to
(\ref{fi}). 
To see this, we must firstly explain the meaning of
the expression $U(z)(A)$. This expression simply corresponds to 
the expansion of the exponential function
 and successive acting
of the commutators on matrix $A$. More
precisely
\begin{eqnarray}\label{Uexp}
U(A)_{ij}=A_{ij}+\frac{i}{2\pi\alpha'}[Y^I,A]_{ij}g_{IJ}
X^J(z)+\nonumber \\
+\frac{1}{2}\left(\frac{i}{2\pi\alpha'}\right)^2 
[Y^I,[Y^K,A]]_{ij}g_{IJ}g_{KL}X^J(z)X^L(z)+\dots \ . \nonumber \\
\end{eqnarray}
The second term in (\ref{Uexp}) for $Y^I$ given in
(\ref{clas}) is equal to
\begin{eqnarray}
\frac{i}{2\pi\alpha'}[Y^I,A]_{ij}g_{IJ}X^J(z)=
\frac{i}{2\pi\alpha'}[y^I_i \delta_{ik}A_{kj}-A_{ik}y_k^I
\delta_{kj}]g_{IJ}X^J(z)
= \nonumber \\
=\frac{i}{2\pi\alpha'}[y^I_i-y^I_j]A_{ij}g_{IJ}X^J(z) \ , \nonumber \\
\end{eqnarray}
where there is no summation over $i,j$. 
 In
the same way the third term in (\ref{Uexp}) gives
\begin{eqnarray}
\frac{1}{2}\left(\frac{i}{2\pi\alpha'}\right)^2
[Y^I,[Y^K,A]]_{ij}g_{IJ}g_{KL}X^J(z)X^L(z)=\nonumber \\
=\frac{1}{2}
\left(\frac{i}{2\pi\alpha'}\right)^2[Y^I_{im},[y^K_m-y^K_j]A_{mj}]g_{IJ}
g_{KL}X^J(z)X^L(z)=
\nonumber \\
=\frac{1}{2}\left(\frac{i}{2\pi\alpha'}\right)^2\left(y^I_m\delta_{im}(y^K_m-y^K_j)A_{mj}-
(y^K_i-y^K_m)A_{im}\delta_{mj}y^I_j\right)g_{IJ}
g_{KL}X^J(z)X^L(z)=
\nonumber \\
=\frac{1}{2}\left(\frac{i}{2\pi\alpha'}\right)^2\left(
y^I_i(y^K_i-y^K_j)A_{ij}-(y^K_i-y^K_j)
y^I_jA_{ij}\right)
 g_{IJ}
g_{KL}X^J(z)X^L(z)=\nonumber \\
=\frac{1}{2}A_{ij}\left(\frac{i}{2\pi\alpha'}\right)^2(y^I_i-y^I_j)g_{IJ}X^J(z)(y^K_i-y^K_j)
g_{KL}X^L(z)=\nonumber \\
=\frac{1}{2}A_{ij}\left(\frac{i}{2\pi\alpha'}(y^I_i-y^I_j)
g_{IJ}X^J(z)\right)^2 \ ,\nonumber \\
\end{eqnarray}
where from the fourth row   there is no summation over $i,j$.
To show that (\ref{nphi}) really corresponds
to (\ref{fi}) for $Y^I$ given in (\ref{clas})  we
use the proof by the mathematical 
induction.
Let us presume that the following 
relation is valid for any $P$
\begin{eqnarray}
\left(\frac{i}{2\pi\alpha'}\right)^P[Y^{I_1},[
Y^{I_2},\dots, [Y^{I_P},A]]]_{ij}
g_{I_1J_1}g_{I_2J_2}\dots g_{I_PJ_P} X^{J_1}(z)
X^{J_2}(z)\dots X^{J_P}(z)=\nonumber \\
=A_{ij}\left
(\frac{i}{2\pi\alpha'}(y^I_i-y^I_j)g_{IJ}X^J(z)\right)^P \ .
\nonumber \\
\end{eqnarray}
Then for $P+1$ we have
\begin{eqnarray}
\left(\frac{i}{2\pi\alpha'}\right)^{P+1}[Y^{I_1},[
Y^{I_2},\dots, [Y^{I_{P+1}},A]]]_{ij}
g_{I_1J_1}g_{I_2J_2}\dots g_{I_{P+1}J_{P+1}}\times 
\nonumber \\ \times
X^{J_1}(z) X^{J_2}(z)\dots X^{J_{P+1}}(z)=\nonumber \\
=\frac{i}{2\pi\alpha'}[Y^K_{ik}A_{kj}
\left(\frac{i}{2\pi\alpha'}(y^I_k-y^I_j)g_{IJ}X^J(z)\right)^P-\nonumber \\
-\left
(\frac{i}{2\pi\alpha'}(y^I_i-y^I_k)g_{IJ}X^J(z)\right)^PA_{ik}
Y^K_{kj}]g_{KL}X^L(z)
= \nonumber \\
=\left(\frac{i}{2\pi\alpha'}\right)^{P+1}\left(
y^K_iA_{ij}\left
((y^I_i-y^I_j)g_{IJ}X^J(z)\right)^P-\right.\nonumber \\
\left. -y^K_jA_{ij}
\left
((y^I_i-y^I_j)g_{IJ}X^J(z)\right)^P\right)g_{KL}X^L(z)
=\nonumber \\
=A_{ij}\left
(\frac{i}{2\pi\alpha'}(y^I_i-y^I_j)g_{IJ}X^J(z)\right)^{P+1} \ , \nonumber \\
\end{eqnarray}
where again  there is no summation over $i,j$ from the fourth row.
Using the previous result we obtain the expression
\begin{equation}
U(A)_{ij}(z)=A_{ij}\exp \left(\frac{i}{2\pi\alpha'}
(y^I_i-y^I_j)g_{IJ}X^J(z)\right) \ 
\end{equation}
without summation over $i,j$.
We then see that (\ref{nphi})
has a correct form of the  matrix valued string field
for the description of the string configuration in
the background (\ref{clas}) of $N$ D-instantons. 

Before we turn to the next example,
 we must certainly find some consistency
conditions which these generalised conformal operators should 
obey. We will proceed as follows. 
Let us start with general  configuration of
$N$ D-instantons described with any $U(N)$ valued
matrices $Y^I$. Then we  require that  the matrix valued string
field $\hat{\Phi}$  should 
obey linearised equation of motion of
the string field theory action. In abelian
case this leads to the requirement that given
state is annihilated by the BRST operator $Q_B$.
It is reasonable to presume that this holds in
nonabelian case as well so we obtain the condition
\begin{equation}\label{lineareq}
Q_B\ket{\hat{\Phi}}=0 \ , \ket{\hat{\Phi}}=
\Phi(0)U(A)(0)\ket{0} \ . 
\end{equation}
We will study consequence of this equation. 
In order to do that we must find an operator product expansion (OPE)
between various matrix valued operators and stress energy
tensor of the open  string theory
\begin{equation}\label{T}
T(z)=-\frac{1}{\alpha'}\partial_z X^{I}(z)\partial_zX^{J}(z)g_{IJ} , \
X^{I}(z)X^{J}(w)=-\frac{1}{2}\alpha'\ln (z-w)g^{IJ} \ ,
I, J=1,\dots ,26 \ . 
\end{equation}
Using (\ref{T}) we can easily calculate 
 OPE between $T(z)$ and $U(0)$. For example,
let us consider the OPE between stress energy
tensor (\ref{T}) and the first two terms in the expansion
of $U(0)$ acting on any $A$ corresponding
to any CFT operator $\Phi$
\begin{eqnarray}
T(z)\frac{i}{2\pi\alpha'}[Y^I,A]g_{IJ}X^J(0)\sim 
\frac{1}{z}\frac{i}{2\pi\alpha'}[Y^I,A]g_{IJ}\partial_zX^J(0) \ , \nonumber \\
T(z)\frac{1}{2}\left(\frac{i}{2\pi\alpha'}\right)^2
[Y^{I_1},[Y^{I_2},A]]g_{I_1J_1}g_{I_2J_2}X^{J_1}(0)
X^{J_2}(0)\sim \nonumber \\ \sim 
\frac{1}{2z}\left(\frac{i}{2\pi\alpha'}\right)^2
[Y^{I_1},[Y^{I_2},A]]g_{I_1J_1}g_{I_2J_2}
\partial_z (X^{J_1}(0)X^{J_2}(0))-\nonumber \\
-\frac{\alpha'}{4z^2}
\left(\frac{i}{2\pi\alpha'}\right)^2[Y^I,[Y^J,A]]g_{IJ} \ .
\nonumber \\
\end{eqnarray}
Generally we have
\begin{eqnarray}\label{genOPE}
T(z)\frac{1}{P!}\left(\frac{i}{2\pi\alpha'}\right)^P
[Y^{I_1},[Y^{I_2},\dots,[Y^{I_P},A]]]g_{I_1J_1}\dots g_{I_PJ_P}
\times \nonumber \\
\times X^{J_1}(0) \dots X^{J_P}(0)
\sim \frac{1}{z}
\frac{1}{P!}\left(\frac{i}{2\pi\alpha'}\right)^P
\sum_{k=1}^P 
[Y^{I_1},[Y^{I_2},\dots,[ Y^{I_k},\dots,[Y^{I_P},A]]]
\times \nonumber \\
\times g_{I_1J_1}\dots
g_{I_kJ_k}\dots g_{I_PJ_P} 
X^{J_1}(0)\dots  X^{J_{k-1}}(0)\partial_z X^{J_k}(0)
X^{J_{k+1}}(0)\dots X^{J_P}(0)- \nonumber \\
-\frac{1}{z^2}\frac{\alpha'}{4P!}
\left(\frac{i}{2\pi\alpha'}\right)^P\sum_{m=1,n=2,m\neq n}^P
g_{IJ}g^{IJ_m}g^{JJ_n}[Y^{I_1},[Y^{I_2},\dots,[Y^{I_m},\dots,
[Y^{I_n},\dots,[Y^{I_P},A]]]\times \nonumber \\
\times g_{I_1J_1}\dots g_{I_PJ_P}
 X^{J_1}(0)\dots X^{J_{m-1}}(0)X^{J_{m+1}}\dots X^{J_{n-1}}
(0)X^{J_{n+1}}(0)\dots X^{J_P}(0) \ . \nonumber \\
\end{eqnarray}
From the  previous  expression we can deduce
that generally there  is not well defined OPE between
stress energy tensor and matrix valued string field.
As a consequence of this fact we cannot define how
such a matrix valued string field transforms under conformal
transformations and hence we cannot define string
field action. 
In fact, in analogy with the abelian case we would like to have
an OPE in the form  
\begin{equation}
T(z)\hat{\Phi} (0) \sim \frac{1}{z^2} h (\hat{\Phi})(0)+\frac{1}{z}\partial_z \hat{\Phi}(0) \ ,
\end{equation}
where $h(\hat{\Phi})$ is a conformal dimension of given field $\hat{\Phi}$. 
 In order to obtain OPE in similar form we demand that
the  background configuration of D-instantons obeys following
rule 
\begin{equation}\label{com}
[Y^I,[Y^J,B]]-[Y^J,[Y^I,B]]=0 , \ \forall B \ , I,J=1,\dots, 26 \ ,
\end{equation}
or equivalently
\begin{equation}\label{eback}
[[Y^I,Y^J],B]=0 \Rightarrow [Y^I,Y^J]=i\theta^{IJ}1_{N\times N} \ 
\end{equation}
that holds of course only in case of infinite dimensional
$U(N)$ 
matrices $Y^I$. We can
expect that this configuration describes higher dimensional
D-brane with noncommutative world-volume.
 In  case of finite dimensional matrices the
only possible solution is
\begin{equation}\label{eback1}
[Y^I,Y^J]=0 \ .
\end{equation}
 Conditions (\ref{eback}),(\ref{eback1})   are
 precisely solutions of the
 equation of motion arising from the low energy action for
$N$ D-instantons.
We have seen the similar result in our previous paper
\cite{KlusonMSFT} where the requirement of the nilpotence
of the matrix valued BRST operator leads to the conclusion
that  the background configuration of D-instantons must obey
equation (\ref{eback}), (\ref{eback1}).

Using (\ref{com}) we can  move  $Y^{I_{m}}$ and then
$Y^{I_n}$ on the left hand side of the second expression
in (\ref{genOPE}). Since we have $P$ possible $I$ and 
$P-1$ $J$ and all appear in the expression in the symmetric way,
the summation in the second expression
 in (\ref{genOPE}) gives  the factor $P(P-1)$ so that
the second term in (\ref{genOPE}) gives
\begin{eqnarray}
-\frac{\alpha'P(P-1)}{4P!z^2}\left(\frac{i}{2\pi\alpha'}\right)^P
g_{IJ}[Y^I,[Y^J,[Y^{I_1},\dots,[Y^{I_{P-2}},A]]]\times 
\nonumber \\
\times 
g_{I_1J_1}
\dots g_{I_{P-2}J_{P-2}}X^{J_2}(0)\dots X^{J_{P-2}}(0)=
\nonumber \\
=-\frac{\alpha'}{4z^2(P-2)!}\left(\frac{i}{2\pi\alpha'}\right)^2
g_{IJ}[Y^I,[Y^J,\left(\frac{i}{2\pi\alpha'}[Y^K,\cdot \ ]g_{KL}
X^L(0)\right)^{P-2}A] \nonumber \\
\end{eqnarray}
and the first equation in (\ref{genOPE}) gives
\begin{eqnarray}
\frac{1}{z}
\frac{1}{P!}\left(\frac{i}{2\pi\alpha'}\right)^P
\sum_{k=1}^P 
[Y^{I_1},[\dots, [Y^{I_k},[\dots,[Y^{I_P},A]]]\times \nonumber \\
\times 
g_{I_1J_1}\dots
g_{I_kJ_k}\dots g_{I_PJ_P}  X^{J_1}(0)\dots  X^{J_{k-1}}(0)\partial_z X^{J_k}(0)
X^{J_{k+1}}(0)\dots X^{J_P}(0)
= \nonumber \\
=\frac{1}{z P!}\partial_z\left(\left(\frac{i}{2\pi\alpha'}\right)
[Y^I,\cdot \ ] g_{IJ}X^J(0)\right)^P(A) \ .\nonumber \\
\end{eqnarray}
Collecting all previous results we obtain the following
 operator product expansion
\begin{eqnarray}\label{TOPE}
T(z)\hat{\Phi}(0)\sim
\frac{1}{z^2}\left(\frac{1}{16\pi^2\alpha'}g_{IJ}[Y^I,[Y^J,
\hat{\Phi}(0)]]+h_{\Phi}\hat{\Phi}(0)\right)
+\frac{1}{z}\partial_z \hat{\Phi}(0) \ ,
\end{eqnarray}
where $h_{\Phi}$ is the conformal dimension of 
the operator $ \Phi(0)$ in (\ref{nphi}). We see that "the conformal
dimension" of $\hat{\Phi}$ is now matrix valued and depends on the
configuration of various D-instantons. 
 Since  the ghost sector does not
depend on the background configuration of 
D-instantons,
 the acting of the BRST operator on $\Phi$ is the
same as in the abelian case. 
When we also use the gauge
\begin{equation}
b_0\ket{\Phi}=0 \ 
\end{equation}
we see that the linearised equation of motion
(\ref{lineareq}) 
leads to condition 
\begin{equation}\label{mass}
\frac{1}{16\pi^2\alpha'}g_{IJ}[Y^I,[Y^J,
\hat{\Phi}(0)]]+h_{\Phi}\hat{\Phi}(0)=0  \ ,
\end{equation}
where (in gauge $b_0\ket{\Phi}=0$)
\begin{equation}
Q_B\ket{\Phi}=h_{\Phi}\ket{\Phi} \ .
\end{equation}
Condition (\ref{mass}) expresses the fact 
that each  component $\hat{\Phi}_{ij}$
of the matrix valued string field
describes state
of the
open string connecting  the i-th
D-instanton with the j-th D-instanton
 which is on the mass shell. 
For example,   for the diagonal background (\ref{clas}) 
the first term in the  bracket in (\ref{TOPE}) gives
\begin{eqnarray}
\frac{1}{16\pi^2\alpha'}g_{IJ}[Y^I,[Y^J,\hat{\Phi}(0)]]_{ij}=\nonumber \\=
\frac{1}{16\pi^2\alpha'}g_{IJ}
[y^I_i\delta_{ik}[Y^J,\hat{\Phi}(0)]_{kj}-[Y^J,\hat{\Phi}(0)]_{ik}\delta_{kj}y^I_j]=
\nonumber \\
=\frac{1}{16\pi^2\alpha'}g_{IJ}(y^I_i-y^I_j)(y^J_i-y^J_j)
\hat{\Phi}_{ij}(0) \ \nonumber \\
\end{eqnarray}
(again no summation over $i,j$). Then  we have 
 a natural result that 
 the conformal dimension of
each component $\hat{\Phi}_{ij}$
is proportional to the distance
between the i-th D-instanton and the j-th
D-instanton.

The OPE between generalised matrix valued string field
 and
the stress energy tensor has also an
 important consequence for the conformal transformation
of given string field and hence 
for the generalised form of the string field action. 
Recall, that a primary vertex operator of conformal dimension $h$
 transforms under $z'=f(z)$ as
\begin{equation}
\mathcal{O}'(z')=\left(\frac{\partial f}{\partial z}\right)^{-h}
\mathcal{O}(z)=\exp\left\{ \left(-h\ln (f'(z)\right)\right\}\mathcal{O}(z) \ .
\end{equation}
From the second form of this description and from the fact that for the
matrix valued string field the first term in
(\ref{TOPE})   acts as a matrix on
given string field we can anticipate following generalised matrix
valued conformal transformation
\begin{equation}\label{confgen}
\hat{\Phi}'(z')=\exp\left(-\ln(f(z))\left[
\frac{1}{16\pi^2\alpha'}g_{IJ}[Y^I,[Y^J,\cdot \ ]]+h_{\Phi}\right]
\right)(\hat{\Phi})(z) \ ,
\end{equation}
where the exponential function should be understood as a matrix
valued function and its action on $\hat{\Phi}(z)$ in form of the Taylor expansion
and where $h_{\Phi}$  is the conformal dimension of the operator $\Phi(z)$.
In particular, for the background (\ref{clas}) we have
\begin{equation}
\hat{\Phi}(z)_{ij}= A_{ij}\exp\left(
\frac{i}{2\pi\alpha'}(y^I_i-y^I_j)g_{IJ}X^J(z)\right) \ .
\end{equation}
In order to determine the behaviour of
this field under conformal transformation,
we must expand  exponential function in
(\ref{confgen}) and let it to act on  $\hat{\Phi}$.
For example, for the
background (\ref{clas}) we obtain 
the result that the
field $\hat{\Phi}_{ij}$ that describes the
state of the string going from the i-th D-instanton
to the  j-th D-instanton transforms under the general
conformal transformation according to the usual rule
\begin{equation}
\hat{\Phi}'(z')_{ij}=\left(\frac{df(z)}{dz}\right)^{-
\frac{1}{16\pi^2\alpha'}(y_i-y_j)^2-h_{\Phi}}
\hat{\Phi}(z)_{ij} \ .
\end{equation}
Now we turn to the second example which is
the background configuration of $N$
 D-instantons (in this case 
 $N\rightarrow \infty$) in the form
\begin{equation}\label{noncommut}
[Y^a,Y^b]=i\theta^{ab}, \ a, b=1,\dots, 2p, \
Y^m=0  , \ m=2p+1,\dots, 26 \ .
\end{equation}
As in the previous case  we begin with (\ref{Uexp}) where
the second term is proportional to
\begin{equation}
i[Y^I,A]g_{IJ}X^J(0) \ .
\end{equation}
 Following \cite{Das,Das1,Das2} we introduce the set of matrices
\begin{equation}
O_k=e^{i\theta^{ij}k_ip_j} , \ p_b=\theta_{bc}Y^c \ ,
\theta_{ac}\theta^{cb}=\delta_a^b \ .
\end{equation}
Then we can write any matrix  as follows
\begin{equation}
A=\int d^{2p}k \exp [i\theta^{ab}k_ap_b]A(k) \ ,
\end{equation}
where $A(k)$ is an ordinary function. 
Then it is easy to see \cite{Das1,Das2} 
\begin{equation}\label{Dasid}
[p_i,O_k]=k_i O_k , \ [p_i,p_j]=-i\theta_{ij} 
\end{equation}
and consequently
\begin{eqnarray}
\frac{2\pi}{4\pi^2\alpha'}[Y^a,A]g_{ab}X^b(0)=
[p_a,A]G^{ab}\tilde{X}_b(0)=
\int d^{2p}kk_aG^{ab}\tilde{X}_b(0) A(k)O_k \  ,  \nonumber \\
G^{ab}=-\frac{1}{4\pi^2\alpha'^2}\theta^{ac}g_{cd}\theta^{db}
\ , \tilde{X}_b(0)=2\pi\alpha'\theta_{bc}X^c(0) \ , \nonumber \\
\left(\frac{i}{2\pi\alpha'}\right)^P[Y^{I_1},[
Y^{I_2},\dots, [Y^{I_P},A]]]
g_{I_1J_1}g_{I_2J_2}\dots g_{I_PJ_P}X^{J_1}(z)
X^{J_2}(z)\dots X^{J_P}(z)=\nonumber \\
=i^P\int d^{2p}kk _{a_1}G^{a_1b_1}\tilde{X}_{b_1}(z)\dots
k_{a_P}G^{a_Pb_P}\tilde{X}_{b_P}(z)
A(k) O_k \ . \nonumber \\
\end{eqnarray}
Then it follows 
\begin{equation}
U(A)(z)=\int d^{2p} k \exp \left(i k \tilde{X}(z)\right) 
A(k)O_k, \ k\tilde{X}=k_aG^{ab}\tilde{X}_b \ 
\end{equation}
and consequently
for any conformal field theory operator $\Phi(z)$
(corresponding to some particular $A$)
we get the matrix valued string field  $\hat{\Phi}$ 
\begin{equation}\label{cftng}
\hat{\Phi}=U(A)\Phi(z)=
\int d^{2p} k \Phi(z) \exp \left(i k\tilde{X}(z)\right)A(k)
O_k \ .
 \end{equation}
We  define generalised matrix valued
vertex  operators the form of which we can deduce from
(\ref{cftng}) 
\begin{equation}\label{vermat}
V(k,\Phi(z))=\Phi(z) \exp \left(i k\tilde{X}(z)\right)
O_k 
\end{equation}
It is important  to include the matrix $O_k$ into the definition
of the matrix valued vertex operator $V$ in order
to stress its matrix nature  since any  correlation
function of these operators contains the trace over
matrix indices. Let us consider two such matrix valued vertex
operators
\begin{equation}
V(k_1,\Phi)(z)=O_{k_1}\Phi(z)e^{ik_1 X(z)} \ ,
V(k_2,\Psi)(z)=O_{k_2}\Psi(z)e^{ik_2 X(z)} \ ,
\end{equation}
which should appear in the calculation of the correlation function
and in particular in the string field theory action. 
Let us calculate the generalised OPE of these two operators
where we include the matrix multiplication.
In fact, the calculation of the OPE is an easy task. Matrix multiplication
affects only the parts containing $O_{k_1}, O_{k_2}$ that gives
\begin{equation}
O_{k_1}O_{k_2}=\exp \left(i\theta^{ij}(k_1+k_2)_ip_j-\frac{1}{2}
i\theta^{ab}k_ak_b\right)=e^{-\frac{i}{2}\theta^{ab}k_ak_b}
O_{k_1+k_2}
\end{equation}
using
\begin{equation}
[i\theta^{ab}k_ap_b,i\theta^{cd}k_cp_d]=
-\theta^{ab}\theta^{cd}k_ak_c(-i\theta_{bd})=-i\theta^{ab}k_ak_b
\end{equation} 
and also 
using the relation
\begin{equation}
e^Ae^B=e^{A+B+\frac{1}{2}[A,B]}
\end{equation}
that is  valid 
for operators whose commutator  is  a pure number. 
 Then we have
\begin{eqnarray}
V(k_1,\Phi)(z)V(k_2,\Psi)(w)= O_{k_1}\Phi(z)e^{ik_1 \tilde{X}(z)} 
O_{k_2}\Psi(w)e^{ik_2 \tilde{X}(w)}\sim
e^{-\frac{i}{2}\theta^{ab}k_ak_b}
O_{k_1+k_2}\times \nonumber \\
\times\left( (z-w)^{\frac{\alpha'}{2}k_{1a}G^{ab}k_{2b}}
\exp\left(i(k_1+k_2)\tilde{X}(w)\right)\Phi(w)\Psi(w)+\dots\right) \nonumber \\
\end{eqnarray}
where dots mean other possible singular terms arising from
the expansion of $e^{ik\tilde{X}(z)}$ and from the
 OPE between $\Phi(z)$ and $\Psi(w)$. We
see that the previous OPE has the same form as
the OPE of the vertex operators in the presence of the
background field $B_{ab}=\left(\frac{1}{\theta}\right)_{ab}$ as
is well known from the seminal paper \cite{WittenNG}. 
It is also important  to stress that thanks to the redefinition
$X^c(z)=\frac{1}{2\pi\alpha'}\theta^{cd}\tilde{X}_d(z)$ the
stress  energy tensor looks like
\begin{eqnarray}
T(z)=-\frac{1}{\alpha'}\partial X(z)^I\partial X(z)^Jg_{IJ}=
\nonumber \\
=-\frac{1}{\alpha'}\left(\frac{1}{4\pi^2\alpha'}
\theta^{ac}\tilde{X}_c(z)\theta^{bd}\tilde{X}_dg_{ab}
\right)-\frac{1}{\alpha'}\partial X^i(z)\partial X^j(z)g_{ij}=\nonumber \\
=-\frac{1}{\alpha'}\partial \tilde{X}_a(z)\partial \tilde{X}_b(z)
G^{ab}-\frac{1}{\alpha'}\partial X^i(z)\partial X^j(z)g_{ij} \ . \nonumber \\
\end{eqnarray}
In other words, the  world-sheet stress energy tensor
is expressed in terms of the open string metric $G_{ab}$ in dimensions labelled
with $\tilde{X}^a,\tilde{X}^b,\dots$ hence the OPE 
between the part of the
stress energy tensor depending on the open string metric
and any matrix valued vertex operator
is a function of the open string quantities  only again
with agreement with \cite{WittenNG}.  

We  should also study the generalised conformal
transformation (\ref{confgen}) 
\begin{equation}\label{confgen2}
V'(k,\Phi,z')=\exp\left(-\ln(f(z))\left[
\frac{1}{16\pi^2\alpha'}g_{IJ}[Y^I,[Y^J,\cdot \ ]]+h_{\Phi}\right]
\right)V(k,\Phi,z) \ .
\end{equation}
In fact, $h_{\Phi}$ is given solely
by the conformal dimension of $\Phi(z)$ and the matrix
multiplication  defined in the exponential function in
(\ref{confgen2}) acts on  $O_k$ only. Then we
can expand the exponential function and use
(\ref{Dasid}). It is easy to see that 
we  obtain the standard conformal transformation of
the vertex operator with the momentum $k_a$
\begin{equation}\label{Tr}
V'(k,\Phi,z')=\exp\left(-\ln(f(z))\left[
\frac{\alpha'}{4}k^2+h_{\Phi}\right]\right)
V(k,\Phi,z)=
\left(\frac{df(z)}{dz}\right)^{-\frac{\alpha'k^2}{4}-h_{\Phi}}
V(k,\Phi,z) \ 
\end{equation}
with $k^2=k_aG^{ab}k_b$.
From this expression  we see that  $V(k,\Phi,z)$ has the
conformal dimension equal to $\alpha'k^2/4+h_{\Phi}$ as we
could expect. We can also calculate the OPE between the
stress energy tensor and matrix valued vertex operators.
Since the OPE between vertex operator and
stress energy tensor  determines the conformal
dimension of given operator and this  is known
from (\ref{Tr}), we do not need work out this OPE
and can determine its form directly from (\ref{Tr}).

{\bf More general example}

 Let us consider the background configuration of
D-instantons in the
form 
\begin{equation}
Y^a= 1_{N\times N}\otimes y^a , \
Y^i=\left(\begin{array}{cccc}
y^i_1\otimes 1 & 0 & \dots & 0 \\
0 & y_2^i \otimes 1 & \dots & 0 \\
\dots & \dots & \dots  & \dots \\
0 & \dots & \dots & y^i_N \otimes 1 \\
\end{array}\right) \ , i=2p+1,\dots, 26 \ ,
\end{equation}
where
\begin{equation}
[y^a,y^b]=i\theta^{ab} \ , a,b=1,\dots, 2p 
\end{equation}
are infinite dimensional
 matrices. It is easy to see
that this configuration obeys (\ref{eback}) 
and hence corresponds to the consistent
background configuration. 
Now we will write any 
 matrix $A$  as follows
\begin{equation}
A=\left(\begin{array}{cccc}
A_{11} & A_{12} & \dots & A_{1N} \\
A_{21} & A_{22} & \dots & A_{2N} \\
\dots & \dots & \dots  & \dots \\
A_{N1} & \dots & \dots & A_{NN} \\ \end{array}\right) \ ,
\end{equation}
where $A_{xy}, \ x, y=1,\dots, N $ are infinite dimensional matrices. 
Let us write any $A_{xy}$ in the form
\begin{equation}
A_{xy}=\int d^{2p}k_{xy}A_{xy}(k_{xy})\exp
\left (i\theta^{ab}k_{axy}p_b\right) \ .
\end{equation}
Now we can write
\begin{eqnarray}
\frac{i}{2\pi\alpha'}[Y^I,A]_{xy}g_{IJ}X^J(z)=
\frac{i}{2\pi\alpha'}[\delta_{xz}\otimes y^a,A_{zy}]g_{ab}X^b(z)+
\nonumber \\
+\frac{i}{2\pi\alpha'}[y^i_{x}\delta_{xz}\otimes 1, A_{zy}]g_{ij}X^j(z)=
\nonumber \\ =
[p_a,A]_{xy}G^{ab}\tilde{X}_b(z)+
\frac{i}{2\pi\alpha'}(y^i_x-y^i_y)g_{ij}X^j(z)A_{xy}= 
\nonumber \\
=i\int d^{2p}k_{xy} \left( k_{xya}G^{ab}\tilde{X}_b(z) 
+\frac{1}{2\pi\alpha'}(y^i_x-y^i_y)g_{ij}X^j(z)\right) A_{xy}(k_{xy})O_{k_{xy}} \ .
 \nonumber \\
\end{eqnarray}
The second term in (\ref{Uexp}) gives
\begin{eqnarray}
\left(\frac{i}{2\pi\alpha'}\right)^2[Y^I,[Y^J,A]]g_{IK}g_{KL}X^K(z)X^L(z)
=\nonumber \\
=i^2\int d^{2p}k_{xy}k _{xya_1}G^{a_1b_1}\tilde{X}_{b_1}(z)
k_{xya_2}G^{a_2b_2}\tilde{X}_{b_2}(z)
A_{xy}(k_{xy}) O_{k_{xy}}+\nonumber \\
+i^2\int d^{2p}k_{xy}A(k_{xy})_{xy}\left(\frac{1}{2\pi\alpha'}(y^i_x-y^i_y)
g_{ij}X^j(z)\right)^2+ \nonumber \\
+2i^2\int d^{2p}k_{xy}O_{k_{xy}} k_{xya}G^{ab}\tilde{X}_b(z)
\frac{1}{2\pi\alpha'}(y^i_x-y^i_y)g_{ij}X^j(z)= \nonumber \\
=i^2\int d^{2p}k_{xy}O_{k_{xy}}A(k_{xy})_{xy}
\left(k_{axy}G^{ab}\tilde{X}_b(z)+\frac{1}{2\pi\alpha'}
(y^i_x-y^i_y)g_{ij}X^j(z)\right)^2 \ . \nonumber \\
\end{eqnarray}
Generally, we have
\begin{eqnarray}
\left(\frac{i}{2\pi\alpha'}\right)^P[Y^{I_1},[
Y^{I_2},\dots, [Y^{I_P},A]]]_{xy}
g_{I_1J_1}g_{I_2J_2}\dots g_{I_PJ_P}X^{J_1}(z)
X^{J_2}(z)\dots X^{J_P}(z)=\nonumber \\
=i^P\int d^{2p}k_{xy}\left(k _{xya}G^{ab}
\tilde{X}_{b}(z)+\frac{1}{2\pi\alpha'}
(y^i_x-y^i_y)g_{ij}X^j(z)\right)^P
A(k_{xy})_{xy} O_{k_{xy}} \ . \nonumber \\
\end{eqnarray}
Using these results we can write the
 generalised matrix valued string
field (\ref{nphi}) in the form
\begin{equation}\label{nphi1}
\hat{\Phi}_{xy}(z)=\int d^{2p}k_{xy}
A(k_{xy})_{xy}\exp\left(ik_{xy}\tilde{X}(z)+
\frac{i}{2\pi\alpha'}(y^i_x-y^i_y)g_{ij}X^j(z)\right)
O_{k_{xy}} \ .
\end{equation}
In summary, we have  obtained the
matrix valued string field for   configuration
of $N$ D2p-branes with the noncommutative world-volume
in dimensions labelled with $x^a, a=1, \dots, 2p$, that
are placed in the different transverse positions labelled
with $y^i_{x}, \ i=2p+1,\dots, 26, \ x=1,\dots, N$.

We hope that  three examples given above 
sufficiently support our proposed form of
the generalised matrix string field (\ref{nphi}).
Then we propose that the string field theory
action for any configuration of 
$N$ D-instantons obeying (\ref{eback}) 
has a form
\begin{equation}\label{actmat}
S=\frac{1}{g_0^2}\tr\left(\frac{1}{2\alpha'}
\left<I\circ \hat{\Phi}(0)Q_B\hat{\Phi}(0)\right>
+\frac{1}{3}\left<f_1\circ \hat{\Phi}(0)f_2\circ
\hat{\Phi}(0)f_3\circ \hat{\Phi}(0)\right>
\right) \ ,
\end{equation}
where  now conformal transformations $
I\circ \hat{\Phi}(0), \
f_i\circ \hat{\Phi}(0), \ i=1,2,3$
are defined by (\ref{confgen}). 
 The precise
study of this action, its particular solutions will
be performed in the forthcoming work.

\section{Conclusion}
In this  paper  we have proposed the matrix valued
form of the string field
that  could be useful for description of
D-instanton configuration using 
 the string field theory action written
in the conformal field  theory language
\cite{Leclair1,Leclair2}.  We have
calculated  OPE of these matrix valued string
fields   with the stress energy tensor.
We have seen that the condition that the
OPE is well defined
 leads to the requirement that the  background
configuration of D-instantons should obey the
equation that can be interpreted as the equation of
motion arising from the low energy matrix theory action.
We have also proposed the generalised
 conformal transformation
of the matrix valued string fields.

As a next step of our research we will study  the
proposed matrix valued string field theory action
(\ref{actmat}). We will also  extend
this approach to the supersymmetric case. 
\\
\\
{\bf Acknowledgement}

We would like to thank Rikard von Unge for
careful reading  the manuscript and many useful
comments. 
This work was supported by the
Czech Ministry of Education under Contract No.
143100006.


\end{document}